\documentclass[slac_one,amsmath]{revtex4}
\usepackage{graphicx}
\usepackage{fancyhdr}
\newcommand{\beq}{\begin{equation}}
\newcommand{\eeq}{\end{equation}}
\newcommand{\beqa}{\begin{eqnarray}}
\newcommand{\eeqa}{\end{eqnarray}}

\begin{document}

\title{
Semi-numerical evaluation of one-loop corrections}
\author{G.Zanderighi} \affiliation{CERN, TH Division, CH-1211, Geneva
  23, Switzerland\\
 Fermilab, P.O. Box 500, 60510 Batavia, IL, US}

\begin{abstract}
  We present a semi-numerical method to compute one-loop corrections
  to processes involving many particles. 
  We treat in detail cases with up to five external legs and massless
  internal propagators, although the method is more general.
\end{abstract}
\maketitle
\thispagestyle{fancy}
\section{INTRODUCTION} 
The importance of next-to-leading order (NLO) corrections is today
well-established.
The benefits of NLO computations include the possibility of reliably
estimating cross-section normalizations, of reducing 
renormalization scale dependences and of understanding the
uncertainties due to the perturbative expansions.
For all searches of new physics at upcoming collider experiments it is
crucial to know the background in good detail, the more so, the
smaller the signal to background ratio is.
At the LHC and at the ILC most processes will involve many particles
in the final state. It is therefore of increasing importance to
predict cross sections for those processes at NLO. 
In QCD all $2\to2$ processes are today known, however a few $2\to3$,
most $2\to 4$ and all $2\to N$ processes with $N >5$ are not yet known
at NLO.

A full $N$ particle NLO calculation requires
i) a tree level (N+1)-particle amplitude;
ii) the NLO N-particle amplitude;
iii) the subtraction terms to regulate the divergences of
 both, i) and ii).
 While point i) has been extensively treated in the literature and
 automatized~\cite{TREE}, and point iii) is also
 well-understood~\cite{SUB}, the bottleneck in N-particle NLO
 calculations remains the complexity of the {\em analytical}
 evaluation of the virtual contribution.
 The aim of this project is therefore to seek a {\em semi-numerical}
 solution to this. Here we report on early steps along this direction.
 For other recent progress in developing algorithms to evaluate
 one-loop integrals see for instance refs.~\cite{LOOPS}.

\section{SEMI-NUMERICAL METHOD}
We first introduce some notation.
We define the $D$ dimensional N-particle M-tensor integral as
\begin{equation}
\label{eq:Iten}
I^{\mu_1 \dots \mu_M}(D; \nu_1,\dots, \nu_N) \equiv
\int \frac{d^D l}{i \pi^{D/2}} 
\frac{l^{\mu_1}\dots l^{\mu_M}}{d_1^{\nu_1} 
\dots d_N^{\nu_N}}\,,\qquad 
d_i \equiv (q_{i}+l)^2\,, \qquad 
q_{i} \equiv\sum_{k=1}^i p_k\,,\qquad 
\sigma \equiv \sum_{i=1}^N \nu_i\,. 
\end{equation}
Notice that we consider here only the case with massless internal
propagators, although the method is more general.

\subsection{The algorithm}
The numerical procedure we used is  simple and is based on
the following few steps~\cite{Ellis:2005zh}
\begin{enumerate}
\vspace{-0.2cm}
\item use a Feynman diagram generator (e.\ g.\ 
  Qgraf~\cite{Nogueira:1991ex}) to generate the amplitude $A$ for a
  specific process at NLO; 
\vspace{-0.2cm}
\item use a symbolic manipulation program (e.\ g.\ 
  Form~\cite{Vermaseren:2000nd}) to write the amplitude as
\begin{equation}
A(p_1,\dots,p_N; \epsilon_1,\dots\epsilon_N; \dots ) = 
\sum_n K_{\mu_1 \dots \mu_N}(p_1,\dots,p_N;
\epsilon_1,\dots\epsilon_N; \dots ) 
\cdot I^{\mu_1 \dots \mu_M}(D; \nu_1,\dots, \nu_N)\>,
\end{equation}
where the kinematic tensor $K$ depends on the particle properties
(momenta $p_i$, polarization $\epsilon_i$, \dots) and is made up of
four-dimensional objects only (i.\ e.\ dependence on the metric {\em
  must} be canceled analytically); \vspace{-0.2cm}
\item use Davydychev reduction formula~\cite{Davydychev:1991va}
\begin{eqnarray} \label{Davyd}
I_{\mu_1\ldots\mu_M}(D;\{\nu_1 \dots,\nu_N\}) =
\sum_{
\substack{
\lambda,\kappa_1,\kappa_2,\ldots, \kappa_N  \ge 0\\
2 \lambda+\sum_i \kappa_i =M}
} &&
\hspace{-.4cm}\left(-{\frac{1}{2}}\right)^{\lambda} \;  
\{ [g]^\lambda [q_1]^{\kappa_1} \ldots 
[q_N]^{\kappa_N} \}_{\mu_1\ldots \mu_M} (\nu_1)_{\kappa_1}
\ldots (\nu_N)_{\kappa_N} \; 
\\
& \times& 
I(D+2(M-\lambda);\nu_1+\kappa_1,\ldots,\nu_N+\kappa_N)\nonumber \,,
\end{eqnarray}
to reduce {\em any} tensor integral to higher dimensional scalar
integrals;
\vspace{-0.2cm}
\item use the basic equation of the integration-by-parts
  method~\cite{IBP},  
\beq
\int\frac{d^Dl}{i \pi^{D/2}}\frac{\partial }{\partial l^\mu }\left(
\frac{\left(\sum_{i=1}^N y_i\right)l^{\mu}
+\left(\sum_{i=1}^N y_iq_i^{\mu}\right)}
{d_1^{\nu_1}d_2^{\nu_2}\cdots d_N^{\nu_N}}\right) = 0\,,
\eeq
to derive a complete set of reduction relations. Here complete means
that any integral is reduced to a linear combination of {\em
  analytically} known scalar integrals.
A sample reduction relation one obtains is for instance 
\begin{equation}
\label{eq:recursion2}
I(D;\{\nu_k\}_{k=1}^N)= 
\frac{1}{(D-1-\sigma)\,B}\left(
I(D-2;\{\nu_k\}_{k=1}^N)
-\sum_{i=1}^Nb_iI(D-2;\{\nu_k-\delta_{ik}\}_{k=1}^N)\right)\,,
\end{equation}
where $S_{ij} \equiv (q_i - q_j)^2$, $b_i \equiv \sum_{j=1}^N
Sij^{-1}$ and $B \equiv \sum_{i=1}^N b_i$.
\end{enumerate}
The method is semi-numerical in the sense that point 1) and 2) are
done analytically, once and for all for a given process, while step 3)
and 4) are repeated numerically for each phase space point.
A key point which makes this method efficient is that a record is kept
of all previously computed scalar integrals, so that each one is
computed only once.

\subsection{Treatment of exceptional phase space points}
Implicit in the numerical use of recursive relations such as
eq.~(\ref{eq:recursion2}) is the assumption that the kinematic matrix
$S_{ij}$ is not singular and that $B$ does not vanish.
If $B$ or $\det(S)$ are exactly zero one obtains a simpler set of
relations~\cite{Giele:2004iy}. This is needed for instance in the
treatment of cases with high $N$, where the particle momenta are not
all linearly independent.
More problematic is the treatment of so called {\em exceptional
  momentum configurations}, where an accidental degeneracy causes $B$
or $\det(S)$ to be very small.
In this case, relations such as eq.~(\ref{eq:recursion2}) are in
principle still valid but become numerically unstable.
A solution to this problem was first suggested in~\cite{Giele:2004ub}.
The idea is to exploit the existence of a small quantity,
parameterizing the closeness to the exceptional phase space
configuration, to define expanded reduction relations.  For instance
if $B \ll 1$ one rewrites eq.~(\ref{eq:recursion2}) as
\begin{equation}
\label{eq:recursion2exp}
I(D;\{\nu_k\}_{k=1}^N) = 
\sum_{i=1}^Nb_iI(D;\{\nu_k-\delta_{ik}\}_{k=1}^N),
+\left(D+1-\sigma\right)\,B\,I(D+2;\{\nu_k\}_{k=1}^N)\,,
\end{equation}
where the first term consists of simpler integrals (lower $D$ and/or
$\sigma$) and the second term consists of more difficult integrals,
which are however suppressed by the small parameter $B$ in front.
In a similar way, if $\det(S)$ is small one determines the
eigenvalue(s) corresponding to a small eigenvector and defines
modified, expanded reduction relations.

\subsection{Higgs plus four partons}
As a first application of the method we considered the yet
uncalculated gluon-gluon fusion amplitude $p_1 +p_2 \to p_3+ p_4+H$ at
NLO in the large $m_t$ limit, where the Higgs couples directly to the
gluons via an effective coupling~\cite{Leff}.
The main mechanisms for Higgs production at hadron colliders are
vector boson fusion (VBF) and gluon gluon fusion (GGF). The former
allows a precise determination of the Higgs properties, in particular
of the Higgs couplings.  
Due to the gluons in the t-channel, GGF is characterized by more QCD
activity in the central region, therefore VBF and GGF can be partly
discriminated with suitable kinematical cuts. %
However, GGF remains the dominant background to VBF. A precise
determination of it is therefore mandatory.  We performed the
numerical, partonic calculation~\cite{Ellis:2005qe}.
In the four quark case we performed also an analytical
calculation~\cite{Ellis:2005qe}, which confirmed the validity of the
numerical results.
In the two quark-two gluon and in the four gluon case we verified the
Ward identities and known identities between gluonic amplitudes
(cyclicity, reflection and dual Ward identity and the decoupling
identity for number of flavours $n_{\rm f}=0$)~\cite{ID}.
Since the method is based on an analytical evaluation of basis
integrals no loss of accuracy is expected. Using double precision the
accuracy of the results is of the order of $10^{-12}$ ($10^{-6}$) for
normal (exceptional) phase space points (a even higher accuracy can be
achieved at the price of including more terms in the expansions).

\section{Work in progress}
The combination of real and virtual results and phase space
integration to obtain full predictions for Higgs + dijet production at
the LHC is currently begin performed.
We are also exploring the possibility of refining the method and of
finding a more efficient numerical algorithm and phase space
integration procedure.
\begin{acknowledgments}
This work is done in collaboration with R. K. Ellis and W. Giele. 
\end{acknowledgments}

\end{document}